# Fano interference in quantum resonances from angle-resolved elastic scattering

Prerna Paliwal[1†], Alexander Blech[2†], Christiane P. Koch[2]*, Edvardas Narevicius[1]*

[1]Department of Chemical and Biological Physics, Weizmann Institute of Science, Rehovot 76100, Israel

[2]Dahlem Center for Complex Quantum Systems and Fachbereich Physik, Freie Universität Berlin, Arnimallee 14, 14195 Berlin, Germany

*Correspondence to E. Narevicius <edvardas.narevicius@weizmann.ac.il>, C. P. Koch <christiane.koch@fu-berlin.de>

[†]These authors contributed equally.

**Asymmetric spectral line shapes are a hallmark of interference of a quasi-bound state with a continuum of states. Such line shapes are well known for multichannel systems, for example, in photoionization or Feshbach resonances in molecular scattering. On the other hand, in resonant single channel scattering, the signature of such interference may disappear due to the orthogonality of partial waves. Here, we show that probing the angular dependence of the cross section allows us to unveil asymmetric Fano profiles also in a single channel shape resonance. We observe a shift in the peak of the resonance profile in the elastic collisions between metastable helium and deuterium molecules with detection angle, in excellent agreement with theoretical predictions from full quantum scattering calculations. Using a model description for the partial wave interference, we can disentangle the resonant and background contributions and extract the relative phase responsible for the characteristic Fano-like profiles from our experimental measurements.**

Fano interference, i.e., the interference of a discrete quantum state with a continuum of states giving rise to asymmetric line shapes, was first observed in experiments exciting rare gas atoms to Rydberg states[1] and later found in measurements across nuclear[2], atomic[1,3,4], molecular[5] as well as solid-state physics[6–8]. It was named after Ugo Fano who provided a theoretical understanding of the subject[9] by showing that the interference between a discrete excited state of an atom with a continuum state sharing the same energy leads to the appearance of asymmetric line shapes in the measured excitation spectra, which can be characterized by a 'shape' parameter[10]. In absence of a non-resonant background, Fano profiles reduce to the more conventional Breit-Wigner profile characterized by Lorentzian peaks[11]. While the Fano profiles were first introduced in atomic physics, the underlying interference mechanism is relevant for a wide variety of physical systems - both quantum[1–8] and classical[12–14].

In the context of scattering physics, a similar type of interference was identified by Feshbach in nuclear scattering[15,16]. A scattering resonance is formed whenever the collision energy approaches a bound state located in a closed channel coupled to the scattering channel.

Interference between the two scattering channels results in asymmetric line shapes that are a universal feature in Feshbach resonances. This scattering phenomenon is not limited to nuclear physics and has become a widely used tool to control interactions e.g. in ultracold atomic gases[17] or between quasiparticles such as polaritons[18]. While in multichannel scattering, the appearance of Fano line shapes in the total (angle-integrated) cross section is almost the rule, the opposite is true for the shape or orbiting resonances. The formation of these resonance states is different, arising from tunneling through a potential barrier or the time delay caused by a transition above the barrier. In such cases, if the resonant and the background contributions belong to the same partial wave, an asymmetric peak forms in the total cross section[19] as reported e.g. in elastic neutron scattering[2]. If, however, the background contribution comes from a different partial wave, then there is no interference term in the total cross section due to orthogonality of the partial waves.

The information lost can be recovered from differential cross section measurements, directly revealing interference between different partial waves. Thus, when resonances are probed by an angle-dependent study such as measuring the backward scattering spectrum[20] or partially integrated cross sections[21,22], it is important to consider the non-resonant contribution from other partial waves. If the scattering amplitude of a background partial wave is comparable to the amplitude of the partial wave dominating the resonance, the presence of the resonance manifests itself in a wide variety of line shapes. The corresponding asymmetric profiles of the resonances have been observed in angle-dependent cross section measurements in nuclear[23] and electron[24–26] scattering, but the connection to Fano interference was not made, leaving the underlying mechanism unidentified. Here, we fill the gap between Fano's theory and angle-resolved scattering measurements, providing a simple and intuitive explanation for asymmetric line shapes in angle-resolved cross sections. We demonstrate that interference is responsible for shifts in the peak position observed by probing an orbiting resonance in angle-resolved energy-dependent cross section measurements for cold elastic collisions between metastable helium and ground-state normal deuterium molecules. In our experiments[22], we use velocity map imaging (VMI)[27] combined with one-photon 'threshold' ionization to image the angular distribution with high resolution. We present our results together with state-of-the-art quantum scattering calculations[22] and extract the angle-dependent background phase and amplitude values by fitting our experimentally obtained cross sections to a simple model based on Fano's intuition.

**Results and Discussion**

A schematic of the measurement setup and a detailed discussion of the experimental methods used for acquiring angle-resolved velocity-map ion images have already been covered in Ref.[22] along with a theoretical analysis based on full quantum mechanical coupled channels calculations. It has also been shown that the interaction between deuterium in its ro-vibrational ground state and metastable helium is mainly isotropic. Thus, the molecule-atom elastic collision considered here can be described by a single channel potential. The only mechanism that could couple scattering channels for He*- $D_2$ is anisotropy. However, since the interaction anisotropy is small compared with the rotational constant of the molecule, the lowest energy degree of freedom rotation cannot be coupled. This means that while the excitation to odd rotational states is forbidden due to parity, the anisotropy in itself is too small to couple excited even rotational states, see also ref.[28]. The partial integrated rate coefficients, obtained by integrating over the backward hemisphere in these images, exhibits two major resonances followed by a minor resonance[22]. The transformation from a Breit-Wigner profile in total cross section to Fano line shapes in angle-resolved cross sections can be best demonstrated for narrow isolated resonances, for example the low energy resonance dominated by $l = 5$, where $l$ is the orbital angular momentum of the colliding pair. But unfortunately, due to finite energy and angular resolution of the experiment, we could not obtain energy and angle-resolved rates for this resonance. So here, we consider the orbiting resonance which is dominated by the $l = 6$ partial wave. In the total cross section calculated theoretically, the energy position of this resonance is determined at 4.8 K by fitting the cross section values in the resonance region to a Breit-Wigner profile.

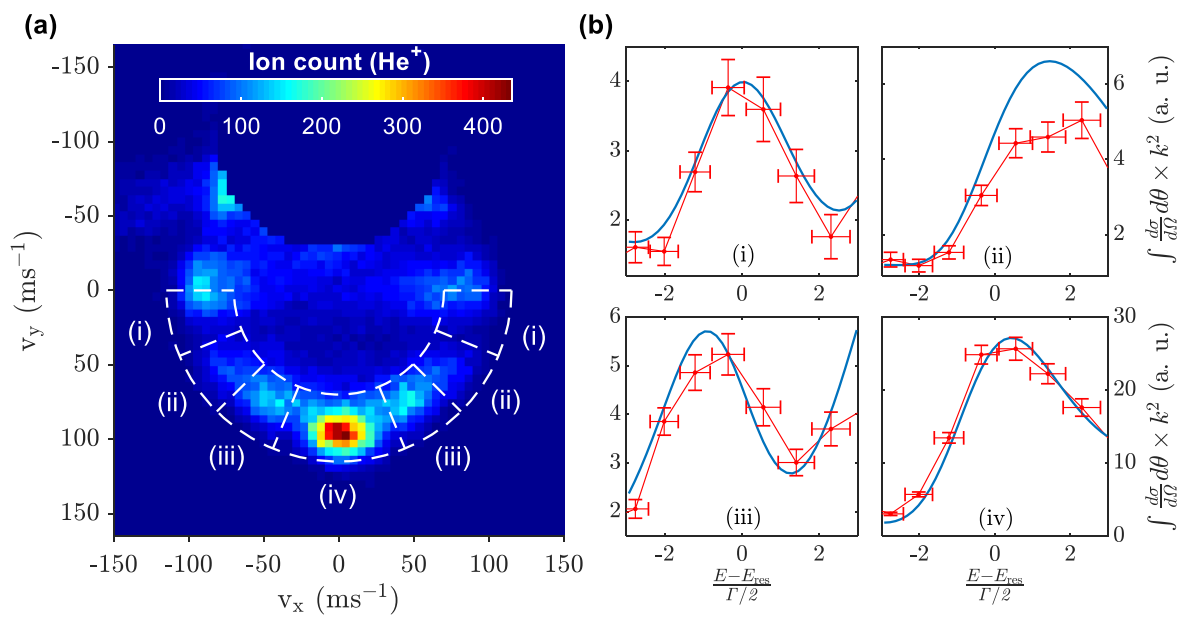

**Fig. 1| Angle-resolved resonance profiles in He*-D₂ elastic scattering.** (a) A VM image is shown at collision energy 4.7 K where different angular sectors are marked as regions (i) to (iv) on the annulus depicted by white-dashed semicircles. The *x*- and *y*-axes represent the velocity of He$^*$ in the centre-of-mass frame. The *y*-axis is the direction of the relative velocity vector and the forward direction ($\theta = 0°$) points up. (b) The experimental (red) and theoretical (blue) angle-resolved energy-dependent cross sections $\times k^2$ (where $k = p/\hbar$ and $p$ is the incident momentum of the colliding pair) are shown in the vicinity of the orbiting resonance dominated by $l = 6$ ($E_{res}$ = 4.8 K, $\Gamma$ = 1.122 K ). The red lines join the experimentally obtained data points. The *x*-axis represents $\frac{E - E_{res}}{\Gamma/2}$ where $E_{res}$ is the resonance energy and $\Gamma$ is the resonance width. The error bars show standard deviation in experimental data points.

A typical VM image obtained from the measurements is shown in Fig. 1 (collision energy 4.7 K), where metastable helium (He*) is detected by single photon 'threshold' ionization at 260 nm. Both unscattered and scattered He* particles are ionized and detected, and they can be distinguished based on scattering angle. In our case, we prepare a cold beam of He* (150 mK) which is crucial for localizing the unscattered particles to a small area on our detector. The information about the particles scattered in the beam direction is masked by the background of unscattered particles. Consequently, the experimental data in the forward section of the image is removed in Fig. 1. The radius of the intense ring visible in this image is proportional to the velocity of scattered He* in the center-of-mass frame. The angular scattering distribution exhibits diffraction oscillations due to interference between different partial waves[22,29]. The pronounced backward scattering visible in the image indicates that the collision energy is close to the resonance energy. The horizontal bands are a result of projecting the particles scattered in a 3D sphere in momentum space onto a 2D detector. Since scattering data in the forward hemisphere is not available, here we determine the angle-dependent cross section from different angular sectors in the backward hemisphere by dividing it into four equal sections represented as (i) to (iv) in Fig. 1 (a). The procedure for obtaining angle-resolved cross section is described in Methods. The experimentally measured cross section determined from these regions in the vicinity of the 4.8 K resonance is shown by red curves in Fig. 1 (b) with error bars representing the standard deviation in the measurements. The x-axis of this graph is the reduced energy $\epsilon$ which measures the collision energy $E$ relative to the position of the resonance $E_{\text{res}}$ in units of its half-width $\Gamma/2$. The width $\Gamma$, which is inversely proportional to the lifetime of the resonance, is determined using the complex absorbing potential method[30]. The theoretically calculated cross sections depicted by solid blue lines have been convoluted with the experimental resolution resulting from the velocity spread of the beams. For all four angular sections, Fig. 1 (b) shows an excellent agreement between the theoretical and experimental cross sections.

The results depicted in Fig. 1 reveal that, depending on the observation angle, the peak of the resonance profile may not necessarily appear at the resonance energy, and may shift to higher or lower energies. In order to rationalize these observations, we apply Fano's reasoning to the differential cross section, see Methods. In our description, we assume that the energy ($E_{\text{res}}$) and width ($\Gamma$) of the resonance remains constant for all angular sections. Parametrizing the resonant contribution $R(E, \theta)$ by a Breit-Wigner amplitude for the dominant partial wave $l_{\text{res}}$, $f_{l_{\text{res}}}(k) = (2l_{\text{res}} + 1)\frac{-1}{(\epsilon+i)}$, where $\epsilon = \frac{E-E_{\text{res}}}{\Gamma/2}$, we find that its interference with the non-resonant background gives rise to Fano line shapes in the differential cross section. While the non-resonant contribution comes from all the background partial waves, the general structure of partial waves governed by Legendre polynomials implies that non-resonant partial waves tend to cancel each other in the backward direction. Thus, their contribution usually is small or negligible and is given by a slowly varying complex scattering amplitude $B(E, \theta)$. To obtain an estimate for the background contribution in the desired angular section, we neglect the energy and angle dependence of background term. Consequently, the background term reduces to a single complex number, with amplitude $A_{bg}$ and phase $\delta_{bg}$. The angle-resolved energy-dependent scattering cross section for the $n$th angular section (with width $\Delta\theta$) is then given by

$$k^2 \int_{\theta_n - \frac{\Delta\theta}{2}}^{\theta_n + \frac{\Delta\theta}{2}} \frac{d\sigma}{d\Omega} d\theta = k^2 \left\langle \frac{d\sigma}{d\Omega} \right\rangle_n = \int_{\theta_n - \frac{\Delta\theta}{2}}^{\theta_n + \frac{\Delta\theta}{2}} |R(E,\theta) + B(E,\theta)|^2 d\theta \qquad (1)$$

$$= \int_{\theta_n - \frac{\Delta\theta}{2}}^{\theta_n + \frac{\Delta\theta}{2}} \left| (2l_{res} + 1) \left( \frac{-1}{\epsilon + i} \right) P_{l_{res}}(\cos\theta) + A_{bg} \exp(i\delta_{bg}) \right|^2 d\theta \qquad (2)$$

where $\left\langle \frac{d\sigma}{d\Omega} \right\rangle_n$ represents the differential cross section integrated over the desired angular region. If the amplitude of $B(E,\theta)$ is negligible, the resonant peak is simply given by the usual Breit-Wigner profile. Conversely, if the amplitude of the background becomes comparable to the resonant amplitude, there will be strong interference near the resonance value. This may lead to peaks, anti-peaks, or a shifted peak, depending on whether the interference is constructive, destructive, or in between.

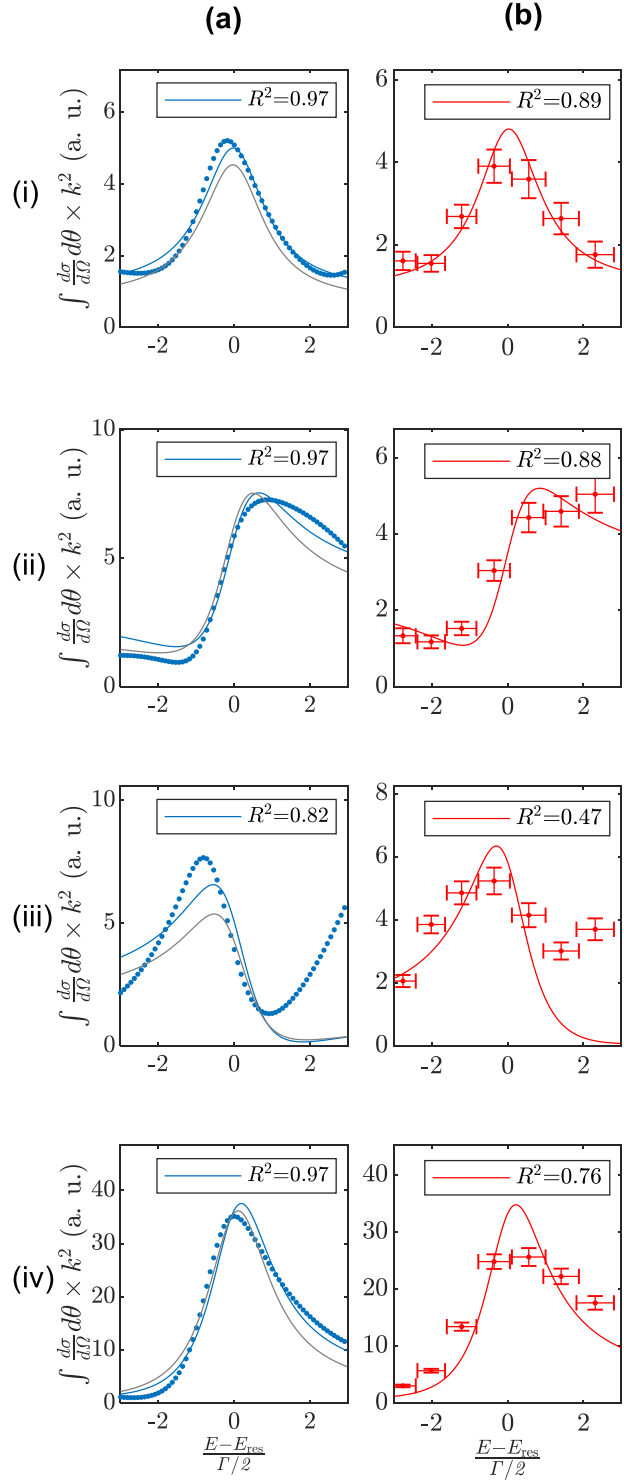

**Fig. 2| Fit of the angle-resolved resonance profiles to a model based on Fano interference.** (a) Theoretically determined angle and energy dependent cross section from coupled channel scattering calculations (dotted blue curves) are fitted to Fano interference model (solid blue lines) in the vicinity of the resonance at 4.8 K. (b) The experimentally measured cross sections (marked in red with error bars, where the error bars indicate standard deviation) are shown along with their corresponding fit to the model (solid red lines). For comparison, we also show grey curves in column (a) generated by approximating the value of background phase and amplitude from quantum scattering calculations and inserting it in Eq. 2. The quality of the fits is assessed by the residual squared ($R^2$) values. The different angular ranges (i) to (iv) are same as labelled in Fig. 1.

In order to estimate the background contribution, we fit the theoretical and experimental cross section values obtained in different angular sections to Eq. (2). The dotted blue curves in Fig. 2, column (a) show the angle-resolved cross sections determined from quantum scattering calculations[22] and the solid blue lines represent the fitted curves. The phase and amplitude values extracted from fitting theoretical data to Eq. (2) are shown in Fig. 3 by blue dots. For comparison, we also obtain an approximate value for $B(E,\theta)$ from quantum scattering calculations by assuming it to be the mid-point value of the calculated $B(E,\theta)$ in the desired energy and angular range. The values thus obtained are shown in Fig. 3 by grey stars. It can be concluded from Fig. 3 that the extracted phase and amplitude values (blue dots) match reasonably well with the values approximated by theory (grey stars), justifying the assumption of energy and angle independence of the background contribution in small angular detection regions (i) to (iv) and in the small energy range defined by the width of resonance. The curves generated by inserting approximated values of $B(E,\theta)$ are also shown in Fig. 2, column (a) by solid grey lines and they match well with the shifts observed in full quantum scattering calculations. This allows us to benchmark our simple model against the full coupled channels quantum scattering calculations. In Fig.2 (b), we fit our experimental data to this model and obtain an experimental estimate for the non-resonant contribution in all these angle-resolved cross sections. The comparatively small value of $R^2$ value obtained for the fits in Fig. 2(a,iii) and Fig. 2(b,iii) can be explained by a strong energy dependence of the background which comprises a contribution from another resonance as explained in Supplementary Fig. 1. The amplitude and phase of the effective background contribution extracted directly from the experimental data are also shown in Fig. 3. It can be seen that the values for relative phase and amplitude of the background extracted from the experimental data match well with the theoretically calculated values. Thus, our model determines the relative phase that is responsible for the measured shifts in the peak of the resonance profile when angle-dependent cross sections are probed. This relative phase is the scattering analogue to the spectral phase that has been recovered before for Fano resonances that are external field induced by attosecond XUV[31,32] or x-ray pulses[33]. In these examples, phase recovery allows for the characterization of electron interaction[31,32] and is a prerequisite for ultrafast x-ray structure determination[33]. In contrast, the relative phase recovered here is not due to an external field but rather acquired intrinsically from the potential governing the molecular scattering. Measurements of the differential elastic scattering cross sections thus allow for recovering the intrinsic phase relation between the interfering states and attest to the quantum coherence in cold collisions that can be harnessed for studying entanglement or controlling reactive scattering at the state-to-state level.

In summary, asymmetric resonance line shapes had so far been observed in nuclear scattering at high collision energies in the range of MeV[23] and electron scattering at energies of a few eV[26]. Here we have reported the observation of such line shapes in molecular scattering with energies of less than 1 meV, made possible by analyzing the angle-dependent cross section of He*-$D_2$ elastic collisions. Our results demonstrate that the peak of the resonance profile may vary from the position of a scattering resonance in an angle-dependent cross section measurement, such as backward or partial rate measurements, within the resonance linewidth. By obtaining the full differential cross section, it is possible to determine the resonance position as well as a relative phase shift with the respect to the effective scattering background. We have further explained how the interference of resonant and background partial waves creates asymmetric, Fano-like profiles in the differential cross section. Our approach generalizes the understanding of asymmetric resonant line shapes in total cross section measurements to the analogous effect arising in angle-resolved cross sections. Applying the intuition of Fano interference, we have developed a model allowing us to extract the effective background contribution directly from the experimentally obtained angle-resolved cross sections. This allows us to disentangle resonance and background contributions purely from the experimental measurements. Whereas in the attosecond experiments the relative phase is generated and controlled by external fields for colliding particles it is an intrinsic property reflecting the interparticle interaction. Our approach should apply to any angle-dependent cross section measurement and will enable the evaluation of background contributions without the need for full-scale quantum scattering calculations. While this work was under review, we have become aware of related work on the angular dependence of molecular photoionization delays in shape resonance[34].

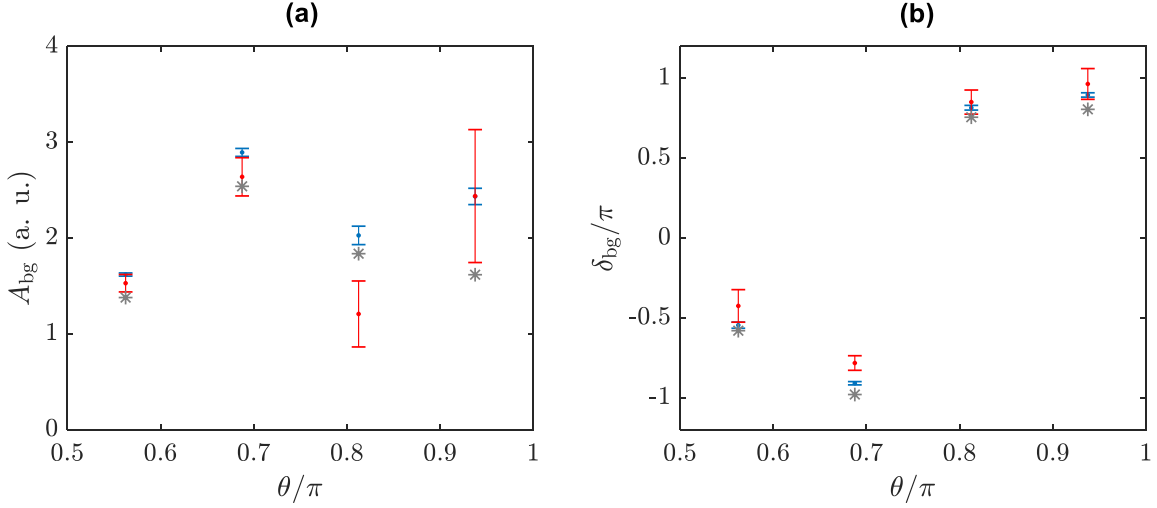

**Fig. 3| Extracted phase and amplitude values for background contribution.** A comparison of the fit parameters – (a) amplitude ($A_{bg}$) and (b) phase ($\delta_{bg}$) – as determined from fitting the experimental (red) and theoretical (blue) data to the model for different angular ranges (i) to (iv) shown in Fig. 1. The grey stars represent the corresponding values calculated by quantum scattering calculations. The error bars correspond to fitting with 95% confidence. The *x*-axis represents the mean value of $\theta$ in different angular sectors labelled as (i) to (iv) in Fig. 1.

**Methods:**

*Experiment:* We merge a pulsed supersonic beam of helium in 2 $^3S_1$ metastable state (He*) with another pulsed supersonic beam of ground state normal deuterium molecules (two-thirds ortho-$D_2$ with *j* = 0 and one-third para-$D_2$ with *j* = 1, where j is the rotational quantum number), using a magnetic guide. The velocity of the He* beam is kept constant at 906 m/s with a standard deviation of 13 m/s throughout the measurement. The velocity of the $D_2$ beam is tuned by changing the temperature of the valve to obtain a mean velocity in the range of 1065–1135 m/s with a standard deviation of about 36 m/s. This results in variable collision energy ranging from 3–6 K with its resolution limited by the spread in the beam velocities. After collisions, we probe scattered He* by single photon threshold ionization at 260 nm to avoid any blurring from electron recoil. The laser polarization direction is along the collision axis and its pulse energy is maintained at 10 microjoules (10 Hz). The He$^+$ ions are then extracted using a velocity map imaging apparatus and imaged on a 2D microchannel plate detector coupled to a phosphor screen. More details of the experiment can be found in Ref.[22]. The energy resolution of the experiment is defined by the velocity spread of the beams, which together with the size of the image determines the angular resolution of the experiment. The calculated cross section shown in Fig. 1 is convoluted with the experimental resolution which is determined to be 0.4 K in this range. The error bars in experimentally measured collision energy result from the uncertainty in the measured velocities of the two supersonic beams.

In order to investigate the structure of the resonance peaks when angle-dependent cross sections are measured, we obtain several velocity map ion images close to the resonance energy. The angle-resolved cross section from the VM images acquired in these experiments is determined by partitioning each image into different angular sectors as shown in Fig. 1. For a given energy, the rate-coefficient is obtained by counting the ions confined within a particular angular region bound by the annulus shown by dashed white circles in Fig. 1 and then divided by the intensities of the reactant beams. The center of the annulus is determined by the peak in the radial distribution of He* ions in the VM image. The width of the annulus is proportional to the radius of the projected image. The resulting rate is further multiplied by $v$ to obtain values proportional to $k^2$ where $v$ and $k$ are the velocity and momentum of the colliding pair. The values thus obtained are corrected to account for the measurement bias that arises due to kinematic effects of the apparatus. This was achieved by multiplying the obtained cross section values with a transfer function generated numerically by dividing the angle-resolved cross section obtained from theory to the angle-resolved cross section obtained by counting the number of ions in the desired angular region bounded by the annulus in the simulated images. The simulated images are created by using theoretically obtained DCS and include all the experimental kinematic effects as described in ref.[22]. This transfer function also brings the experimentally measured values in arbitrary units to an absolute scale.

*Origin of Fano line shapes:* Here, we describe the origin of Fano-profiles in the differential cross section at a specific angle $\theta$, which reads

$$k^2 \frac{d\sigma}{d\Omega} = |\sum_{l=0}^{\infty} f_l(k) \, P_l(\cos\theta)|^2 \tag{3}$$

where $f_l = (2l+1)\sin\delta_l \, e^{i\delta_l}$ and $\delta_l$ is the energy-dependent scattering phase of partial wave $l$.

We assume that there is a single resonance carried by the partial wave $l_{res}$, which we parametrize by the Breit-Wigner form, $\sin^2\delta_l = \frac{1}{1+\epsilon^2}$. Following Fano's reasoning[10], we partition the total scattering amplitude into a resonant and a background contribution,

$$k^2 \frac{d\sigma}{d\Omega} = |R(E,\theta) + B(E,\theta)|^2 \tag{4}$$

with $R(E,\theta) = f_{l_{res}}(k) \, P_{l_{res}}(\cos\theta)$ and $B(E,\theta) = \sum_{l=0, l\neq l_{res}}^{\infty} f_l(k) \, P_l(\cos\theta)$.

By inserting the explicit form of $f_l$, Eq. (4) can be further written as

$$k^2 \frac{d\sigma}{d\Omega} = \left| A_{l_{res}} \sin\delta_{l_{res}} \, e^{i\delta_{l_{res}}} + A_{bg} e^{i\delta_{bg}} \right|^2 \tag{5}$$

where we have defined $A_l = (2l+1) \, P_l(\cos\theta)$,

$$A_{bg} = \sqrt{\sum_{l=0, l\neq l_{res}}^{\infty} A_l^2 \sin^2\delta_l + \sum_{l<l'; l,l'\neq l_{res}} 2 A_l A_{l'} \sin\delta_l \sin\delta_{l'} \cos(\delta_l - \delta_{l'})}$$

$$\delta_{bg} = \arctan\left( \frac{\sum_{l=0, l\neq l_{res}}^{\infty} A_l \sin^2\delta_l}{\sum_{l=0, l\neq l_{res}}^{\infty} A_l \sin\delta_l \cos\delta_l} \right)$$

Following the lines of Refs.[4,35], we match the background-subtracted Fano profile with the mixing terms in the differential cross section to obtain

$$k^2 \frac{d\sigma}{d\Omega} = \sigma_{res} + d\sigma_{bg} + \sigma_0 \frac{q^2 + 2q\epsilon - 1}{1+\epsilon^2} \tag{6}$$

This equation is equivalent to Eq. (5) with the following choice of $q$, $\sigma_0$, $\sigma_{res}$ and $d\sigma_{bg}$,

$q \equiv \cot\left(\delta_{bg} - \frac{\pi}{4}\right)$

$\sigma_0 \equiv \frac{2 A_{l_{res}}(\theta) A_{bg}(\theta)}{1+q^2}$

$\sigma_{res} \equiv A_{l_{res}}^2 \sin^2\delta_{l_{res}} = \frac{A_{l_{res}}^2}{1+\epsilon^2}$

$d\sigma_{bg} \equiv A_{bg} \approx$ constant.

Eq. (6) implies that, for a fixed angle $\theta$, the differential cross section can be written as a combination of a Lorentzian peak, given by $\sigma_{res}$, and a potentially asymmetric Fano line shape.

The strength and the shape of the Fano profile thus depend on the angle-dependent magnitude and phase of the background scattering amplitude.

In order to directly compare with our experimental results, we integrate the differential cross section over the angular regions as defined in Eq. (1) in the main text. In Eq. (2), we approximate $B(E,\theta)$ as constant close to the resonance energy and evaluate an average value of the background amplitude in the desired energy and angular range by fitting the experimentally obtained cross section to Eq. (2). This provides a simple way to estimate the background scattering amplitude directly from experimental data.

**Data availability:**
The source data files for all figures are available at the public repository. (https://doi.org/10.5281/zenodo.5665901).

**Acknowledgements:**
We acknowledge financial support from the European Research Council and the Israel Science Foundation. Correspondence and requests for materials should be addressed to E.N. and C.P.K.

**Author Contributions:**
E.N. and C.P.K. conceived and supervised the project. P.P. performed the experiment. A.B. and P.P. did the theoretical calculations. All authors participated in the discussions, came up with the model and commented on the manuscript.

**References:**

1. Beutler, H. Über Absorptionsserien von Argon, Krypton und Xenon zu Termen zwischen den beiden Ionisierungsgrenzen 2P03/2 und 2P01/2. *Z. für Phys.* **93**, 177–196 (1935).

2. Adair, R. K., Bockelman, C. K. & Peterson, R. E. Experimental corroboration of the theory of neutron resonance scattering. *Phys. Rev.* **76**, 308 (1949).

3. Madden, R. P. & Codling, K. New autoionizing atomic energy levels in He, Ne, and Ar. *Phys. Rev. Lett.* **10**, 516–518 (1963).

4. Ott, C. et al. Lorentz meets Fano in spectral line shapes: A universal phase and its laser control. *Science* **340**, 716–720 (2013).

5. Linn, S. H., Tzeng, W. B., Brom, J. M. & Ng, C. Y. Molecular beam photoionization study of HgBr2 and HgI2. *J. Chem. Phys.* **78**, 50–61 (1983).

6. Limonov, M. F., Rybin, M. V., Poddubny, A. N. & Kivshar, Y. S. Fano resonances in photonics. *Nat. Photonics* **11**, 543–554 (2017).

7. Miroshnichenko, A. E., Flach, S. & Kivshar, Y. S. Fano resonances in nanoscale structures. *Rev. Mod. Phys.* **82**, 2257–2298 (2010).

8. Luk'Yanchuk, B. *et al.* The Fano resonance in plasmonic nanostructures and metamaterials. *Nat. Mater.* **9**, 707–715 (2010).

9. Fano, U. Sullo spettro di assorbimento dei gas nobili presso il limite dello spettro d'arco. *Nuovo Cim.* **12**, 154–161 (1935).

10. Fano, U. Effects of configuration interaction on intensities and phase shifts. *Phys. Rev.* **124**, 1866–1878 (1961).

11. Breit, G. & Wigner, E. Capture of slow neutrons. *Phys. Rev.* **49**, 519 (1936).

12. Joe, Y. S., Satanin, A. M. & Kim, C. S. Classical analogy of Fano resonances. *Phys. Scr.* **74**, 259–266 (2006).

13. Satpathy, S., Roy, A. & Mohapatra, A. Fano interference in classical oscillators. *Eur. J. Phys.* **33**, 863–871 (2012).

14. Iizawa, M., Kosugi, S., Koike, F. & Azuma, Y. The quantum and classical Fano parameter q. *Phys. Scr.* **96**, 055401 (2021).

15. Feshbach, H. Unified theory of nuclear reactions. *Rev. Mod. Phys.* **36**, 1076–1078 (1964).

16. Feshbach, H. A unified theory of nuclear reactions. II. *Ann. Phys.* **19**, 287–313 (1962).

17. Chin, C., Grimm, R., Julienne, P. & Tiesinga, E. Feshbach resonances in ultracold gases. *Rev. Mod. Phys.* **82**, 1225–1286 (2010).

18. Takemura, N., Trebaol, S., Wouters, M., Portella-Oberli, M. T. & Deveaud, B. Polaritonic Feshbach resonance. *Nat. Phys.* **10**, 500–504 (2014).

19. Taylor, J. R. *Scattering Theory: The Quantum Theory on Nonrelativistic collisions* (John Wiley & Sons, 1972).

20. Ren, Z. *et al.* Probing the resonance potential in the F atom reaction with hydrogen deuteride with spectroscopic accuracy. *Proc. Natl. Acad. Sci. U. S. A.* **105**, 12662–12666 (2008).

21. Vogels, S. N. *et al.* Scattering resonances in bimolecular collisions between NO radicals and $H_2$ challenge the theoretical gold standard. *Nat. Chem.* **10**, 435–440 (2018).

22. Paliwal, P. *et al.* Determining the nature of quantum resonances by probing elastic and reactive scattering in cold collisions. *Nat. Chem.* **13**, 94–98 (2021).

23. Jungmann, C. R. *et al.* Resonance structure of 32S + n from transmission and differential elastic scattering experiments. *Nucl. Phys.* **A386**, 287–307 (1982).

24. Warner, C. D., King, G. C., Hammond, P. & Slevin, J. Resonance structure in elastic scattering of electrons from atomic hydrogen. *J. Phys. B: Atom. Mol. Phys.* **19**, 3297–3303 (1986).

25. Buckman, S. J. & Clark, C. W. Atomic negative-ion resonances. *Rev. Mod. Phys.* **66**, 539–655 (1994).

26. Regeta, K. *et al.* Resonance effects in elastic cross sections for electron scattering on pyrimidine: Experiment and theory. *J. Chem. Phys.* **144**, 024301 (2016).

27. Eppink, A. T. J. B. & Parker, D. H. Velocity map imaging of ions and electrons using electrostatic lenses: Application in photoelectron and photofragment ion imaging of molecular oxygen. *Rev. Sci. Instrum.* **68**, 3477–3484 (1997).

28. Klein, A. *et al.* Directly probing anisotropy in atom-molecule collisions through quantum scattering resonances. *Nat. Phys.* **13**, 35–38 (2017).

29. Vogels, S. N. *et al.* Imaging resonances in low-energy NO-He inelastic collisions. *Science.* **350**, 787–790 (2015).

30. Moiseyev, N. *Non-Hermitian Quantum Mechanics*. (Cambridge Univ. Press, Cambridge, 2011).

31. Kaldun, A. *et al.* Extracting phase and amplitude modifications of laser-coupled Fano resonances. *Phys. Rev. Lett.* **112**, 103001 (2014).

32. Kotur, M. *et al.* Spectral phase measurement of a Fano resonance using tunable attosecond pulses. *Nat. Commun.* **7**, 10566 (2016).

33. Heeg, K. P. *et al.* Interferometric phase detection at x-ray energies via Fano resonance control. *Phys. Rev. Lett.* **114**, 207401 (2015).

34. Holzmeier, F. *et al.* Influence of Shape Resonances on the Angular Dependence of Molecular Photoionization Delays. *arXiv:2107.09915* (2021).

35. Caselli, N. *et al.* Generalized Fano lineshapes reveal exceptional points in photonic molecules. *Nat. Commun.* **9**, 396 (2018).

# Supplementary information

**Anomalous line shapes observed in Fig. 2 (a) and Fig. 2 (b) of the main text for the angular section (iii):** Here, we show why the rise in cross section at higher energies in Fig. 2 (iii) cannot be explained by our model. The main assumption in the derivation of our model are isolated resonance interacting with an energy independent background. When we have overlapping contributions from two or more resonances, the observed line shape cannot be explained by this simple model. In the case of angular section (iii), besides $l = 6$ resonance, another resonance dominated by $l = 7$ partial wave starts contributing significantly.

The Supplementary Fig. 1 shows the angle-resolved energy-dependent cross section obtained when the contribution from partial wave $l = 7$ is omitted and we observe that the rise of the cross section at higher energies vanishes for the (iii) angular section. Therefore, the increase in cross section observed in Fig. 2 (iii) of the main text at higher energies is caused by the rising edge of the resonance peak dominated by $l = 7$ [ref.[1]] At these energies, we thus have contributions from two different resonances, dominated by $l = 6$ and 7, respectively, which is not included in our model.

**Supplementary References:**

1. Paliwal, P. *et al.* Determining the nature of quantum resonances by probing elastic and reactive scattering in cold collisions. *Nat. Chem.* **13**, 94–98 (2021).

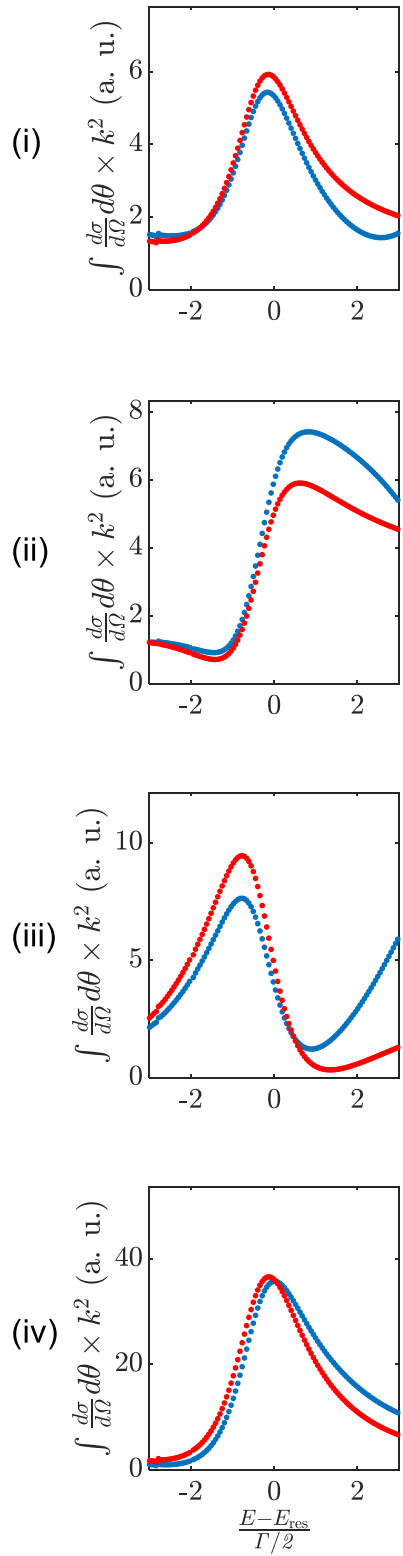

**Supplementary Fig. 1| Resonant contribution from $l = 7$ in the vicinity of energy region dominated by $l = 6$.** The theoretical angle-dependent cross section is shown including the contributions from all partial waves (blue) and omitting the contribution from partial wave $l = 7$ (red) to the cross section for the resonance at 4.8 K. The rising tail of the cross section observed in column (b), row (iii) in Fig. 2 of the main text disappears, indicating that it is caused by the higher energy resonance dominated by $l = 7$.